# Measuring the superhump period of the dwarf nova RX J1715.6+6856

Jeremy Shears, Ian Miller and Richard Sabo

## Abstract

We report unfiltered CCD observations of the first confirmed superoutburst of the dwarf nova RX J1715.6+6856 in August 2009. At quiescence the star was magnitude 18.3 (CCD, clear). The outburst amplitude was at least 2.4 magnitudes and it lasted at least 6 days, although the first part of the outburst was probably missed. Analysis of the light curve revealed superhumps with peak-to-peak amplitude of 0.1 magnitude, thereby showing it to be a member of the SU UMa-family. The mean superhump period was $P_{sh}$ = 0.07086(78) d with a superhump period excess of ε = 0.038 and an estimated mass ratio q = 0.167. In the final stages of the outburst, as the star approached quiescence, the superhumps disappeared to be replaced by a modulation corresponding to the orbital period. The star was regularly monitored between August 2007 and September 2009 revealing a total of 12 outbursts, with an outburst frequency of approximately once per month.

## Introduction

RX J1715.6+6856 was first identified as a dwarf nova by Pretorius *et al.* in the ROSAT North Ecliptic Pole X-ray survey and confirmed via optical spectroscopy which showed features typical of a dwarf nova in quiescence [1]. The magnitude was estimated from the spectroscopic data as V=18.3(2). They determined the orbital period as $P_{orb}$ = 0.06828 d (1.639 h) from Hα radial velocity measurements. Such $P_{orb}$ places RX J1715.6+6856 well below the so-called period gap in the orbital distribution of dwarf novae which occurs between about 2 and 3h [2]. The majority of dwarf novae below the period gap are members of the SU UMa family which exhibit superoutbursts from time to time and the likely SU UMa identity was pointed out by Kato [3]. In general, superoutbursts in SU UMa systems last several times longer than normal outbursts, may be up to a magnitude brighter and the light curve is characterised by superhumps: modulations in the light curve which are a few percent longer than the orbital period. In this paper we report photometry of the first confirmed superoutburst in August 2009. The star is located at RA 17 15 41.50 Dec +68 56 32.0 (J2000).

## Outburst light curve

The outburst was first detected by IM on 2009 Aug 20.887 in an unfiltered (clear, "C") CCD image at C = 15.8 [4]. We conducted time resolved photometry according to the observation log in Table 1, using the instrumentation in Table 2. Differential aperture photometry was performed, after flat fielding and dark subtracting the images, against the sequence from BAA chart P201107 [5]. An image of the star in outburst is shown in Figure 1.

The light curve of the outburst is shown in Figure 2. The beginning of the outburst was not well constrained as the previous observation was on 2009 Aug 7.930 when the star was not in outburst (C > 17.8, IM) and no further observations in the intervening period exist in the AAVSO International Database [6]. On discovery night and the subsequent night, we assume the star was in the plateau phase, which for most SU UMa systems lasts several days. Three nights later the star had already entered the rapid decline phase. During our final time resolved photometry observations, 6 nights after detection, the star was already approaching quiescence. We monitored the star on 8 nights during quiescence after the superoutburst and found a mean magnitude of C=18.3 (range 18.2 to 18.4), which corresponds to the quiescence magnitude reported by Pretorius *et al.* [1]. Thus the outburst lasted at least 6 days and had an amplitude of at least 2.6 magnitudes, although since the beginning of the outburst was missed these are likely to be underestimates.

## Measurement of the superhump period

We plot expanded views of the time series photometry in Figure 3, drawn to the same scale. We observed that there was considerable scatter in the data which appeared to be greater than the photometric noise associated with each data point and which may represent the flickering commonly seen in dwarf novae. To reduce this effect, we plot the average of two data points in Figure 3. Superhumps having a peak-to-peak amplitude of 0.1 magnitude were detected during the first night of photometry (Figure 3a). Three nights later, when the decline had begun, superhumps of similar amplitude were also present (Figure 3b). The following night, however, the modulations were less clear especially towards the beginning of the photometry series (Figure 3c) and, as we shall see below, probably do not represent superhumps. On the final night, when the star was nearly at quiescence (Figure 3d), no regular modulations were apparent by visual inspection of the light curve, although the brightness was varying considerably.

To study the superhump behaviour, we first extracted the times of each sufficiently well-defined superhump maximum during the first two nights on which photometry was conducted (JD 2455065 and 2455068) by fitting a quadratic function to the individual light curves. Times of 6 superhump maxima were found and are listed in Table 3. An analysis of the times of maximum allowed us to obtain the following linear superhump maximum ephemeris:

$$HJD_{max} = 2455065.3772(65) + 0.07086(78) \times E \qquad \text{Equation 1}$$

This gives the mean superhump period $P_{sh}$ = 0.07086(78) d. The observed minus calculated (O–C) residuals for the superhump maxima relative to the ephemeris are shown in Figure 4.

To confirm our measurements of $P_{sh}$, we carried out a period analysis of the data from the first two nights (JD 2455065 and 2455068) using the Lomb-Scargle algorithm in the Peranso software package [7], having subtracted the linear trend from the data. This gave the power spectrum in Figure 5, which has its highest peak at a period of 0.07082(37) d and which we interpret as $P_{sh}$, plus a multiplicity of 1 c/d aliases. This value is consistent with our earlier measurement from the times of superhump maxima. The superhump period error estimate is derived using the Schwarzenberg-Czerny method [8]. Several other statistical algorithms in Peranso gave the same value of $P_{sh}$. A phase diagram of the data from the plateau phase, folded on $P_{sh}$, is shown in Figure 6. This exhibits the typical profile of superhumps in which the rise to superhump maximum is faster that the decline. Given the highly aliased power spectrum we also investigated whether the two peaks above and below the highest peak could actually be the true $P_{sh}$. We did this by plotting phase diagrams of the data folded on the periods of these higher and lower peaks (data not shown). A visual comparison of the resulting phase diagrams, indicated that the best fit was obtained with the data folded on $P_{sh}$ = 0.07082 d, which supports our supposition that the highest peak was due to $P_{sh}$.

While the $P_{sh}$ obtained from the two methods (superhump maximum analysis and period analysis) gave consistent results, it is not unusual for the time of maxima analysis to result in a more accurate method of tracking periodic waves in dwarf novae as it is less troubled by changes in amplitude than period analysis techniques. We note that this analysis is based only on 6 superhump timings, although this is mitigated by the fact that they were obtained over a baseline of 3 days. Hence we adopt our mean value of $P_{sh}$ from the time of maxima analysis as $P_{sh}$ = 0.07086(78) d, even though the formal error on the measurement is greater. Removing $P_{sh}$ from the power spectrum yielded only very weak signals, none of which were related to $P_{sh}$ or $P_{orb}$.

We also carried out a Lomb-Scargle analysis on the de-trended data from the last two nights (JD 2455069 and 2455070) and the resulting power spectrum is shown in Figure 7. The strongest signal was at 0.06944(90) d, or 1.667(22) h, plus its 1 c/d aliases. Such value is close to $P_{orb}$ = 0.06828 d (1.639 h) reported by Pretorius *et al.* [1], although they give no error estimate, hence we suggest it is the orbital signal.

**Other outbursts of RX J1715.6+6856**

In the interval between 2007 Aug 20, when the authors began observing this star, and 2009 Sep 21, a total of 12 outbursts of RX J1715.6+6856 have been detected, including the one discussed in this paper. These outbursts are listed in Table 4, along with the interval between them. It is likely that further outbursts have been missed due to incomplete coverage; we searched the AAVSO International Database, but no other observations are listed apart from those of the authors. The mean outburst interval is 68 days (standard deviation 42 days) and the median is 46 days. Four of the outbursts were separated by less than 40 days and the minimum interval is 29 days, which suggests that an outburst might occur about once a month. In fact the final outburst noted in Table 4 occurred only 30 days after the superoutburst was detected; it attained C=16.9 at maximum and lasted about 2 days. Unfortunately there are insufficient data on amplitude and duration to conclude with certainty which of the other outbursts were normal outbursts or superoutbursts. However, we note that the majority of these outbursts do appear to be low amplitude, often less than 1 magnitude, and are therefore likely to be normal outbursts.

**Discussion**

Taking our mean superhump period of the 2009 outburst, $P_{sh}$ = 0.07086(78) d, and the orbital period from reference 1, $P_{orb}$ = 0.06828 d, we calculate the superhump period excess ε = 0.038. Such value is consistent with other SU UMa systems of similar orbital period [2]. If we assume that RX J1715.6+6856 has a white dwarf of ~0.75 solar masses, as is typical for SU UMa systems, then we can estimate the secondary to primary mass ratio, q, from the empirical relationship ε = 0.18*q + 0.29*$q^2$ [9] as q = 0.137.

The observed 2.4 magnitude outburst amplitude of RX J1715.6+6856 is at the lower end of the distribution of superoutburst amplitudes for the vast majority of SU UMa systems. However, since it is likely that we missed the early part of the outburst, it might have been slightly larger. Another unusual feature is the low amplitude (only 1 to 2 magnitudes or less above quiescence), frequent, and (judging by the last outburst in Table 4) short-lived normal outbursts. Similar behaviour has been observed in the SU UMa systems V1316 Cyg and V452 Cas. V1316 Cyg had a modest 2.4 magnitude superoutburst [10] and faint outbursts of ~1.4 magnitude lasting 1-2 days occurring approximately every 10 days [11]. There has been some debate [11] as to whether these low amplitude brightening events in V1316 Cyg are in fact "normal" outbursts, aborted outbursts or whether they are events caused by flares on the secondary resulting in a short-lived burst of mass transfer [12]. V452 Cas also has low amplitude superoutbursts (3.2 magnitude) and rather small (~1 magnitude) short-lived (<3 days) and frequent (~1 per month) normal outbursts [13, 14].

We encourage further observations of RX J1715.6+6856 with the aim of establishing the nature of the outbursts and interval between normal outbursts and superoutbursts. Additional time resolved photometry during a future superoutburst would be useful to investigate the superhump evolution in more detail, especially if the earlier stage of the superoutburst is included.

**Conclusions**

Analysis of the superoutburst of RX J1715.6+6856 has confirmed that this is a member of the SU UMa family of dwarf novae. The amplitude was at least 2.4 magnitudes. We measured the mean superhump period as $P_{sh}$ = 0.07086(78) d with a superhump period excess of ε = 0.038. Long term monitoring of the star has revealed 12 outbursts over a 25 month period.

**Acknowledgements**

The authors gratefully acknowledge access to the AAVSO International Database, the use of SIMBAD database, operated at CDS, Strasbourg, France and the NASA/Smithsonian Astrophysics Data System. We also thank our referees, Dr. Tim Naylor and Mr. Roger Pickard for their helpful comments which have improved the paper.

**Addresses:**

JS: "Pemberton", School Lane, Bunbury, Tarporley, Cheshire, CW6 9NR, UK [bunburyobservatory@hotmail.com]
IM: Furzehill House, Ilston, Swansea, SA2 7LE, UK [furzehillobservatory@hotmail.com]
RS : 2336 Trailcrest Dr., Bozeman, MT 59718, USA [richard@theglobal.net]

| Start time | Duration (h) | Observer |
|---|---|---|
| 2455065.363 | 3.58 | Miller |
| 2455068.349 | 5.57 | Shears |
| 2455068.425 | 1.61 | Miller |
| 2455068.647 | 3.19 | Sabo |
| 2455069.691 | 6.48 | Sabo |
| 2455070.646 | 7.68 | Sabo |

**Table 1: Log of time-series observations**

| Observer | Telescope | CCD (unfiltered) |
|---|---|---|
| Shears | 0.28 m SCT | Starlight Xpress SXVF-H9 |
| Miller | 0.35 m SCT | Starlight Xpress SXVF-H16 |
| Sabo | 0.43 m reflector | SBIG STL-1001 |

**Table 2: Equipment used**

| Superhump cycle no. | Time of maximum HJD | Error (d) |
|---|---|---|
| 0 | 2455065.3750 | 0.0020 |
| 1 | 2455065.4507 | 0.0018 |
| 2 | 2455065.5191 | 0.0025 |
| 43 | 2455068.4197 | 0.0016 |
| 44 | 2455068.4885 | 0.0019 |
| 45 | 2455068.5731 | 0.0027 |

**Table 3: Times of superhump maximum**

| Discovery JD | ΔT (d) | Discovery UT | | | Magnitude at max(C) | Observer |
|---|---|---|---|---|---|---|
| 2454351 |     | 2007 | Sep | 17 | 16.1 | Shears |
| 2454442 | 91  | 2007 | Dec | 7  | 16.8 | Shears |
| 2454471 | 29  | 2008 | Jan | 5  | 16.3 | Shears |
| 2454571 | 100 | 2008 | Apr | 15 | 17.7 | Miller |
| 2454605 | 34  | 2008 | May | 19 | 16.9 | Miller |
| 2454651 | 46  | 2008 | Jul | 3  | 17.8 | Miller |
| 2454690 | 39  | 2008 | Aug | 11 | 17.4 | Shears |
| 2454756 | 66  | 2008 | Oct | 16 | 17.6 | Shears |
| 2454909 | 153 | 2009 | Mar | 18 | 16.7 | Shears |
| 2455026 | 117 | 2009 | Jul | 14 | 16.5 | Miller |
| 2455064 | 38  | 2009 | Aug | 20 | 15.8 | Miller |
| 2455094 | 30  | 2009 | Sep | 20 | 16.9 | Miller |

**Table 4: Outbursts of RX J1715.6+6856 between 2007 Aug 20 and 2009 Sep 21**

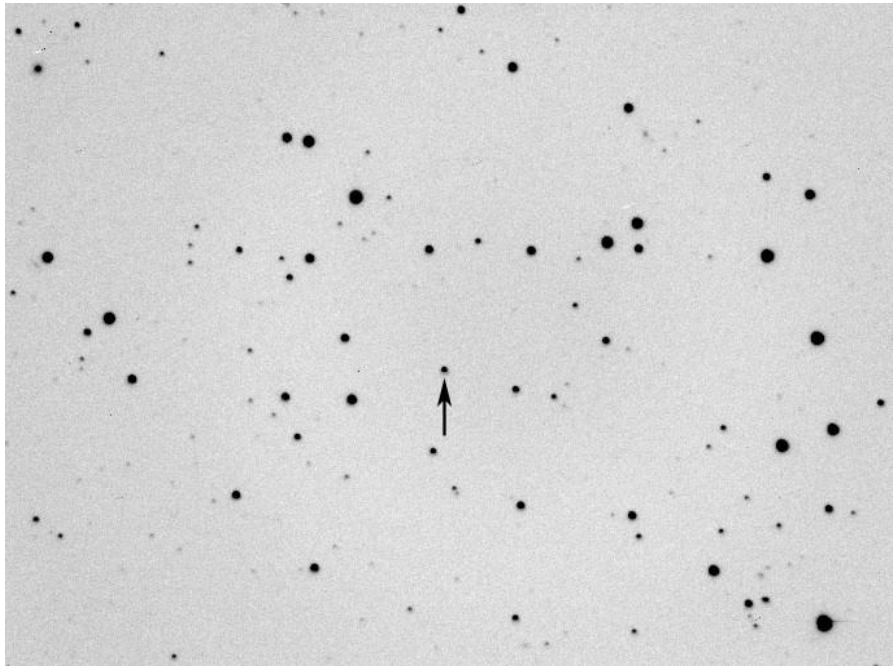

**Figure 1: RX J1715.6+6856 in outburst, 2009 Aug 24.996
Image 12 min by 16 min, N at top E to left**
*(Jeremy Shears)*

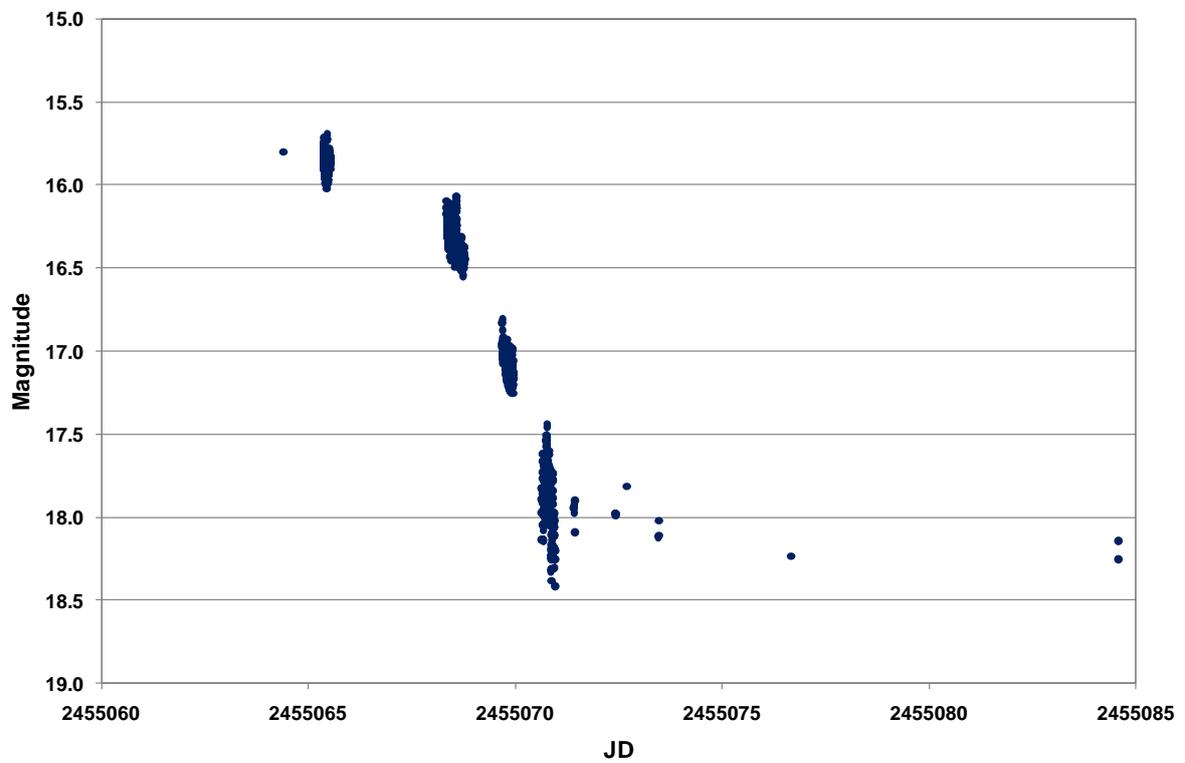

**Figure 2: Outburst light curve**

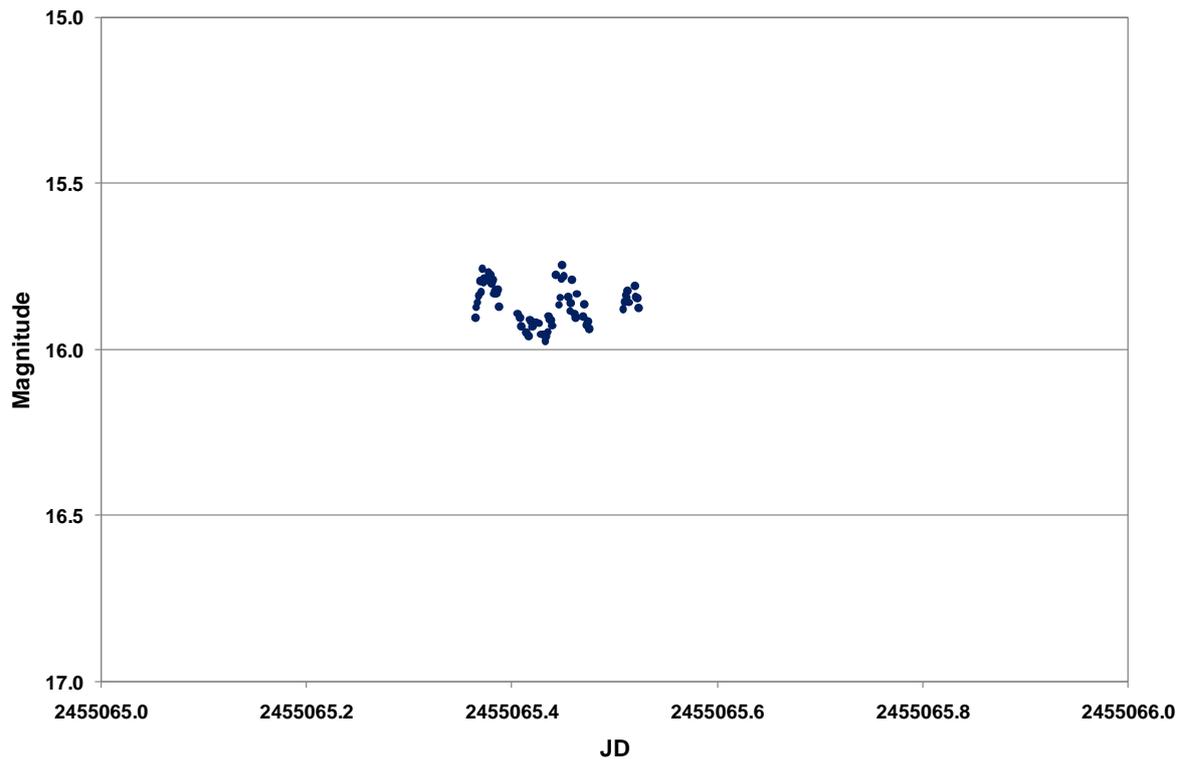

**(a)**

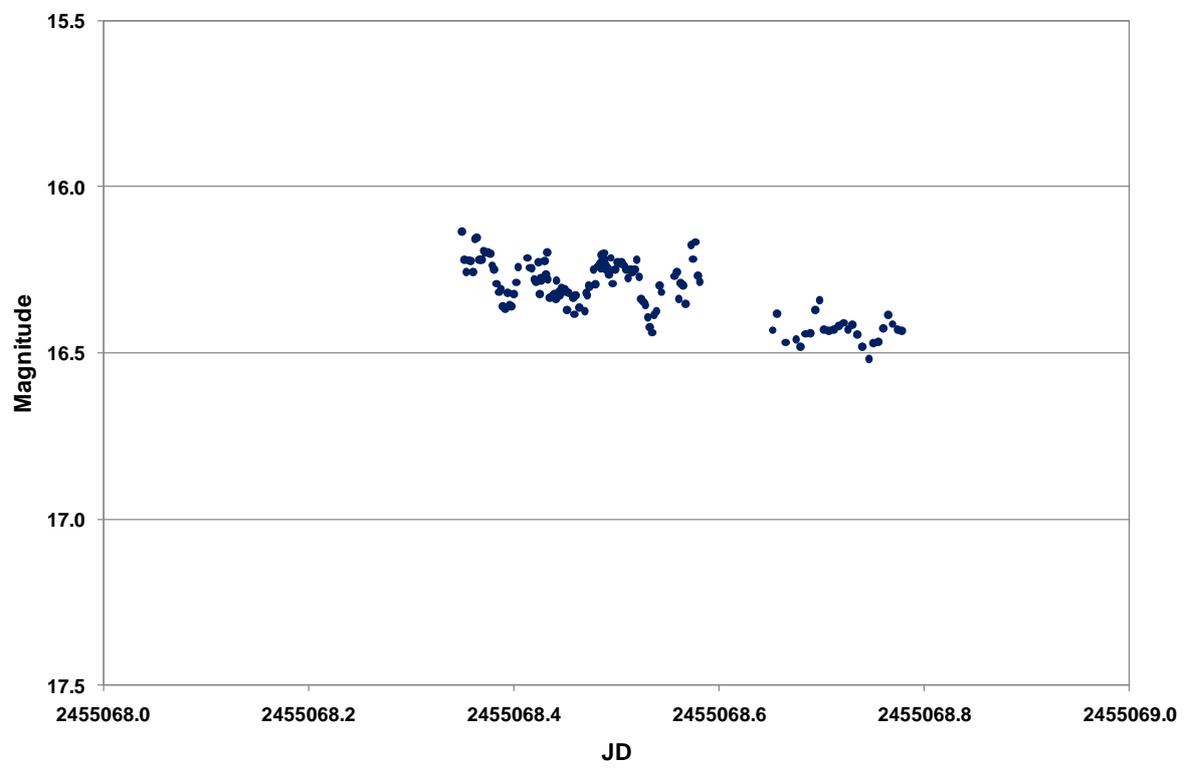

**(b)**

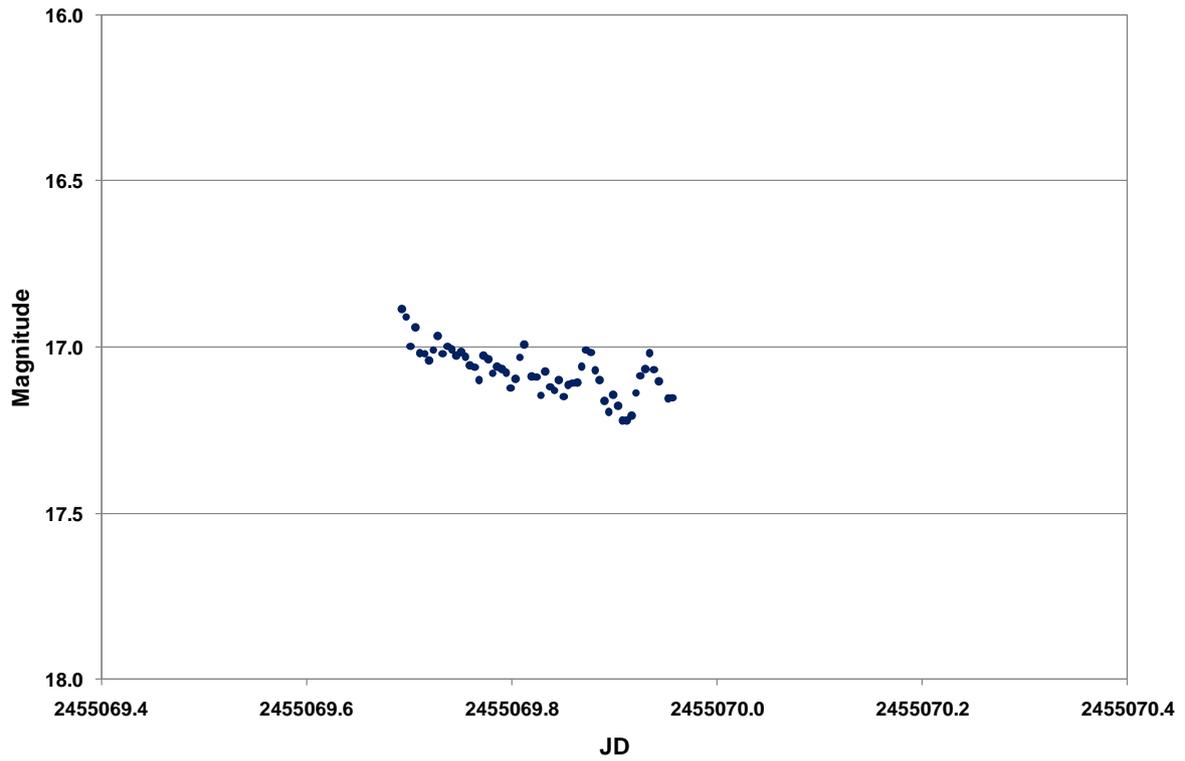

(c)

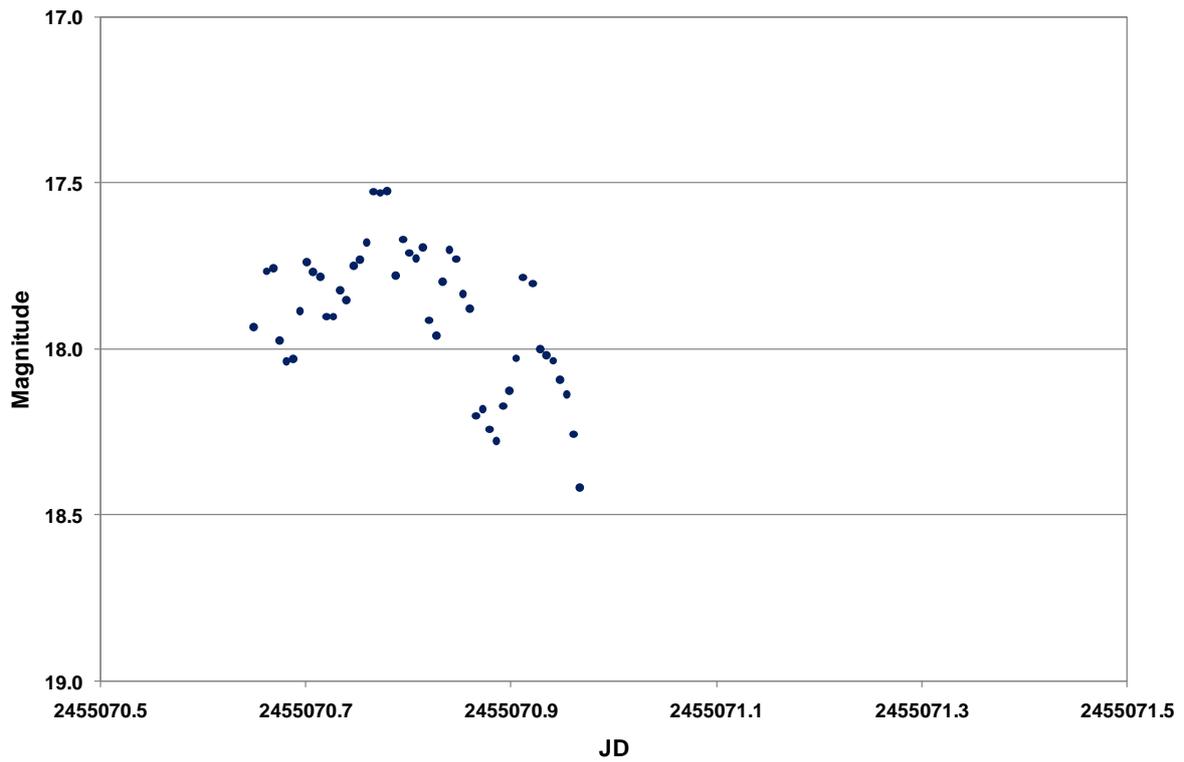

(d)

**Figure 3: Time series photometry during the superoutburst**
*Note: panels (a) and (b) of this Figure appear on the previous page*

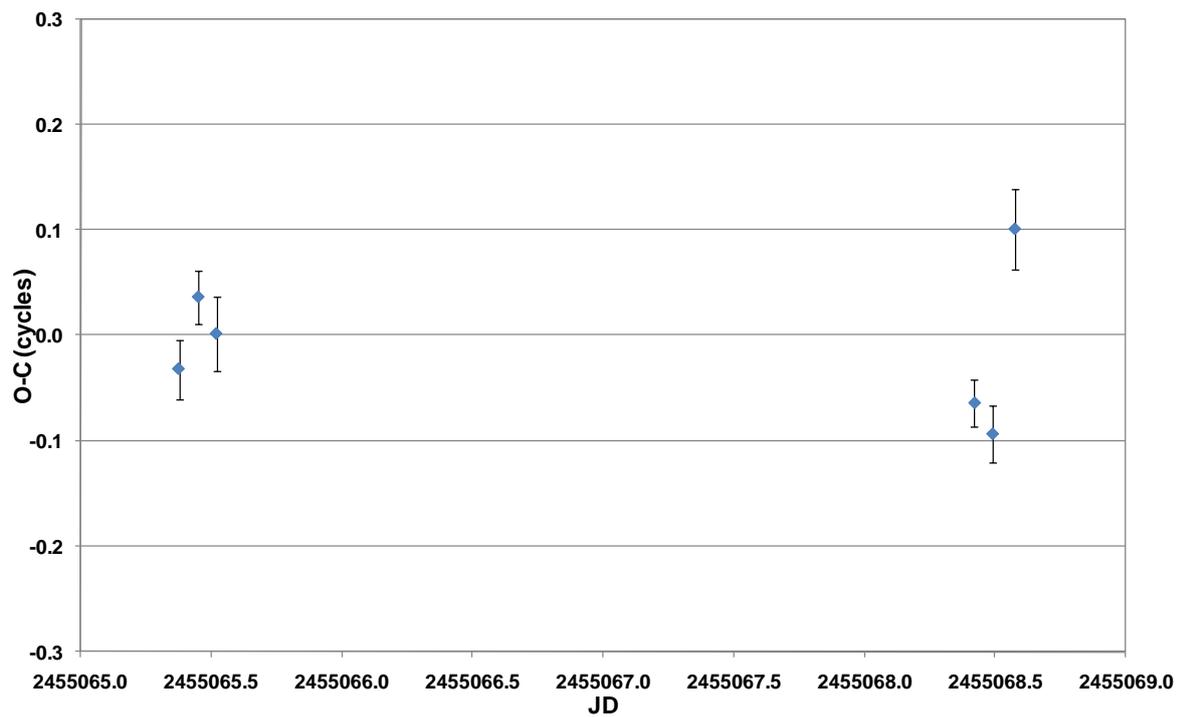

**Figure 4: O-C diagram, data from JD 2455065 and 2455068**

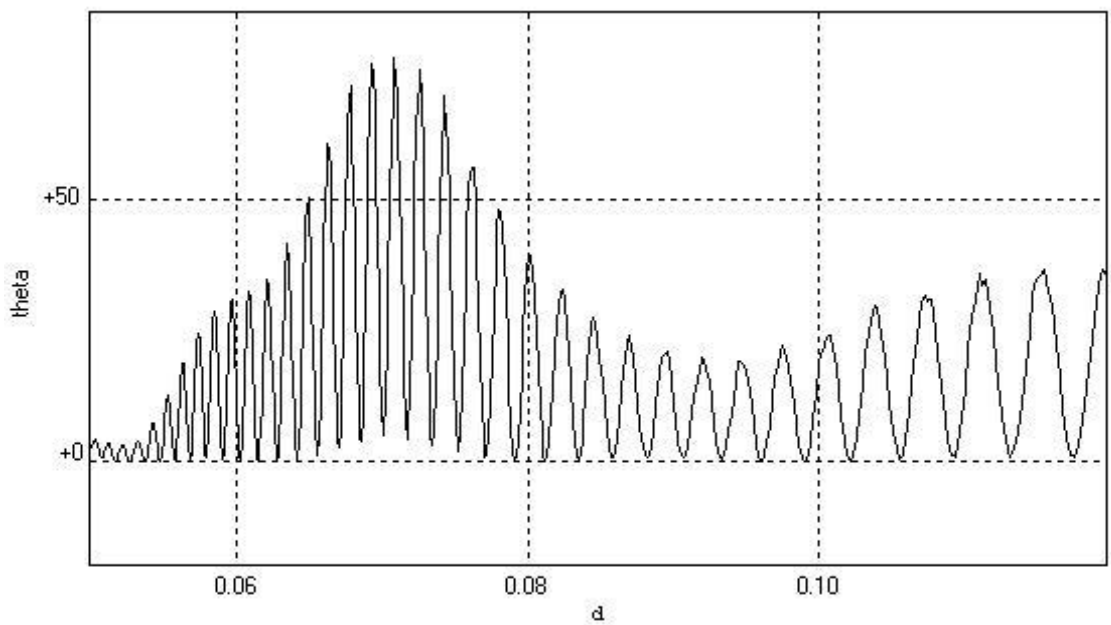

**Figure 5: Power spectrum of combined time-series data from JD 2455065 and 2455068**

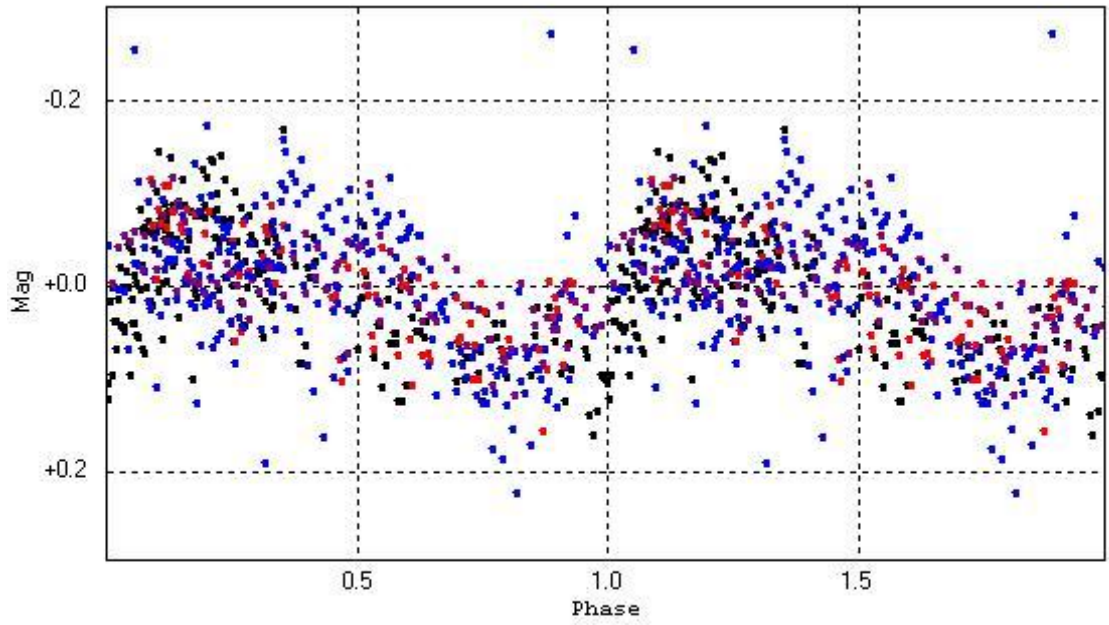

**Figure 6: Phase diagram of data from JD 2455065 and 2455068 folded on P$_{sh}$**

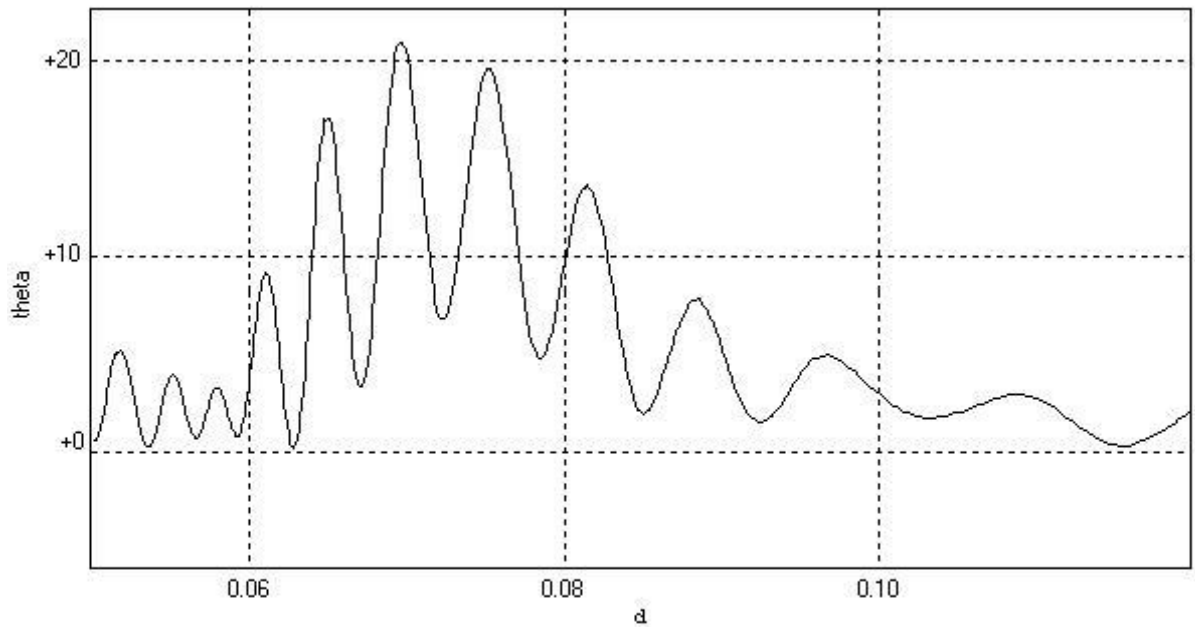

**Figure 7: Power spectrum of combined time-series data from JD 2455069 and 2455070**